\documentclass[]{mn2e}

\usepackage{epsfig} % to have figures in the .eps format
\usepackage{graphicx}
\usepackage{amsmath}
\usepackage[comma]{natbib}
\usepackage{subfigure}
\usepackage[T1]{fontenc}
\usepackage{aecompl}
\usepackage{array}
\usepackage{booktabs}

\begin{document}


\title[Quenching the Main Sequence]{Quenching star formation: insights from the local main sequence}

\author[S. Leslie et al.]{S. K. Leslie$^{1}$\thanks{E-mail:
sarah.leslie@anu.edu.au} , L. J. Kewley$^{1}$,  D. B. Sanders$^2$, N. Lee$^2$\\
$^{1}$Research
School of Astronomy \& Astrophysics, Australian National University, Cotter Road, Weston, ACT 2611, Australia\\
$^{2}$Institute for Astronomy, 2680 Woodlawn Dr., Honolulu, HI, 96822, USA }
\maketitle

\begin{abstract}
The so-called star-forming main sequence of galaxies is the apparent tight relationship between the star formation rate and stellar mass of a galaxy. Many studies exclude galaxies which are not strictly `star forming' from the main sequence, because they do not lie on the same tight relation. Using local galaxies in the Sloan Digital Sky Survey we have classified galaxies according to their emission line ratios, and studied their location on the star formation rate - stellar mass plane. We find that galaxies form a sequence from the `blue cloud' galaxies which are actively forming stars, through a combination of composite, Seyfert, and low-ionization nuclear emission-line region galaxies, ending as `red-and-dead' galaxies. The sequence supports an evolutionary pathway for galaxies in which star formation quenching by active galactic nuclei (AGN) plays a key role.
\end{abstract}


\section{Introduction}
Galaxies can be separated into two distinct classes: those that are actively forming stars and are blue in colour, and those that have no appreciable star formation and are red in colour. Current consensus is that galaxies start their lives in the active `blue' category, and then evolve to be ``red-and-dead'' (e.g. \citealt{goncalves12,moustakas13}). Evolution between these two classes must involve processes which quench the star formation, and quench it relatively quickly (on the order of 1 Gyr), because galaxies with intermediate properties, located in the `green valley', are not often seen \citep{faber07}. The dearth of \textit{observed} green valley galaxies could also be explained if they become obscured, or if their number density is underestimated due to the short variability/duty cycle time-scales of black hole accretion \citep{hickox14}.
A popular candidate quenching mechanism is negative feedback from active galactic nuclei (AGN), because AGN host galaxies tend to have green or red optical colours (e.g. \citealt{nandra07}).

Blue star-forming galaxies form a tight ($\sim$ 0.2 dex intrinsic scatter; \citealt{speagle14}) sequence between star formation rate (SFR) and stellar mass ($M_*$) often referred to as the galaxy `main sequence' of star formation \citep{noeske07, daddi07, elbaz07}. The main sequence (MS) is generally characterized by a single power law, often expressed as 
\begin{equation} \log(\text{SFR})=\beta \log (M_*) +\alpha \end{equation}
with typical values for the slope $\beta =0.7-1.0$. \cite{speagle14} found that the slope, $\beta$, and normalization, or intercept, $\alpha$, increase and decrease respectively, as a function of the age of the Universe $t$ (in Gyr):
\begin{eqnarray}
\beta &=&(0.84 \pm 0.02 - 0.026 \pm 0.003 \times t),\\
\alpha &=&(-6.51 \pm 0.24 + 0.11 \pm 0.03 \times t).
\end{eqnarray}
SFR at 10$^{10}$ M$_\odot$ evolves from $\sim$10 to $\sim$1 M$_\odot$ yr$^{-1}$ from z$\sim$1 to the present day.
The existence of the MS implies that star formation in the majority of galaxies is governed by quasi-steady processes, and may imply that only a small fraction of local galaxies undergo more chaotic processes such as major mergers \citep{lotz11}. Galaxies with quasi-exponentially increasing SFRs and stellar masses at high redshift will form an MS relation with $M_*\propto$ SFR \citep{dave08}. 
The fitted parameters of the MS relation reveal even more about the nature of star formation and stellar mass build up: the slope of the MS is related to the low mass slope of the mass function of star-forming galaxies \citep{peng10}, and the evolution of the MS normalization is thought to result from evolution in gas densities with redshift \citep{whitaker12,magdis12,speagle14}.

Different studies report inconsistent slopes and shapes of the MS. For example, studies by \citet{drory08,karim11, whitaker14} \& \cite{lee15} suggest that the MS requires a more complex, stellar mass dependent fit. \cite{karim11} \& \cite{lee15} report a flattening of the MS at stellar masses $M_* > 10^{10}M_\odot$. On the other hand, after including both star-forming and quiescent galaxies from the Sloan Digital Sky Survey (SDSS; \citealt{york00}), \cite{renzini15} found no deviation from a simple power-law MS relation in local galaxies, which suggests that the deviations might originate from inconsistent definitions of star-forming galaxies and/or selection effects. However, the analysis of \cite{renzini15} excluded galaxies exhibiting AGN activity, because it has been historically difficult to reliably measure their underlying SFRs.

Studies which include AGN in colour-mass diagrams have shown that AGN host galaxies populate the red-sequence or the transitional green valley\citep{nandra07, schawinski07, smolcic09, alatalo14b}. The inclusion of AGN in colour-mass diagrams thus allowed the role AGN feedback in the quenching the star formation of early type galaxies to be inferred (e.g. \citealt{schawinski07}).

In this Letter we use the SDSS to investigate the effects of including all spectral classes of galaxies, in the main sequence analysis. We will show that AGN host galaxies also populate intermediate regions on the $M_*$ - SFR plane. 

\section{Data Sample}
The data used in this study comes from the SDSS data release 7 \citep{abazajian09}, with all properties calculated by the MPA/JHU group\footnote{http://www.mpa-garching.mpg.de/SDSS/DR7/}. We only use data for galaxies with redshifts greater than z $>$ 0.04 to ensure that the SDSS fibre covers at least 30\% (median coverage of 38\%) of the typical galaxy to avoid aperture effects \citep{kewley05}. 
We also constrain our sample to z$<$0.1 to avoid incompleteness and significant evolutionary effects \citep{kewley06}.
This redshift restriction results in a sample of 273,274 galaxies with MPA/JHU derived emission line fluxes, SFRs and stellar masses. We further restrict our sample by applying a signal to noise limit of 3 on the H$\alpha$ emission line, resulting in a sample of 202,708 galaxies.
We have also retrieved reddening corrected rest-frame photometry for these galaxies from the SDSS DR 7 data base. We use the magnitudes resulting from fitting either a de Vaucouleurs profile or a pure exponential profile, which have been corrected for galactic extinction following the determinations of \cite{schlegel98}.

We use the SFRs from the MPA/JHU catalogue, which were calculated based on the methods of \cite{brinchmann04}, with modifications to the methods for non-SF galaxies. For galaxies classified as SF, the SFR is calculated by the MPA/JHU team using the H$\alpha$ calibration of \cite{kennicutt98}\footnote{Note that using the Kroupa IMF, \cite{kennicutt12} present a H$\alpha$ SFR which is 0.17 dex lower than the calibration in \cite{kennicutt98}.}. They use the Balmer decrement and method of \cite{charlot00} to correct for extinction and assume a \cite{kroupa} initial mass function (IMF). For galaxies falling outside of the SF category, the MPA group calculated their SFRs estimated based on the strength of the D4000 break (defined in \citealt{balogh98}). In \cite{brinchmann04}, an empirical relationship was derived between H$\alpha$ SFR and D4000, for SF galaxies. An updated version of the resulting relationship is used to calculate SFRs for galaxies in which the H$\alpha$ flux may not be an accurate indicator of SFR (e.g. in the presence of AGNs, or quiescent galaxies with low S/N emission lines). Unlike in \cite{brinchmann04}, the aperture corrections were made for SFRs by the MPA/JHU team by fitting the photometry of the outer regions of the galaxies \citep{salim07}. 

Stellar masses are calculated by the MPA/JHU team using a fit to photometry. The stellar masses in the MPA/JHU catalogue, have been found to be consistent with other estimates (see \citealt{taylor11,chang15}).
In the following analysis we use the median SFR and M$_*$ from the probability distributions for each galaxy, however using the mean does not change our results. 

\subsection{Classification}
We classify each galaxy according to the \cite{kewley06} scheme which makes use of the emission line ratios [NII]/H$\alpha$, [SII]/H$\alpha$, [OI]/H$\alpha$, and [OIII]/H$\beta$. These emission line ratios were proposed in diagnostic diagrams by \cite{baldwin81} and \cite{veilleux87} to classify the dominant energy source in an emission-line galaxy. We refer to the optical classification diagrams as the `standard optical diagnostic diagrams'.

We find that out of the 202,708 galaxies in our sample, 60.4\% are classified as SF, 12.2\% are classified as composite, 4.1\% are classified as Seyferts , 6.5\% are classified as LINERs (low-ionization nuclear emission-line regions) and 17.0\% are classified as ambiguous- i.e. they do not fall into the same category on all the three diagnostic diagrams. The fraction of ambiguous galaxies becomes smaller if signal-to-noise limits are applied to the forbidden lines. Most galaxies classified as ambiguous lie close to the Seyfert 2-LINER or composite-AGN boundaries. We expect some of these ambiguous galaxies to have the properties of composite, Seyfert 2, or LINER galaxies. 

\section{The `Galaxy Main Sequence' including AGN hosts}
\begin{figure*}
\centering
\includegraphics[width=0.95\linewidth]{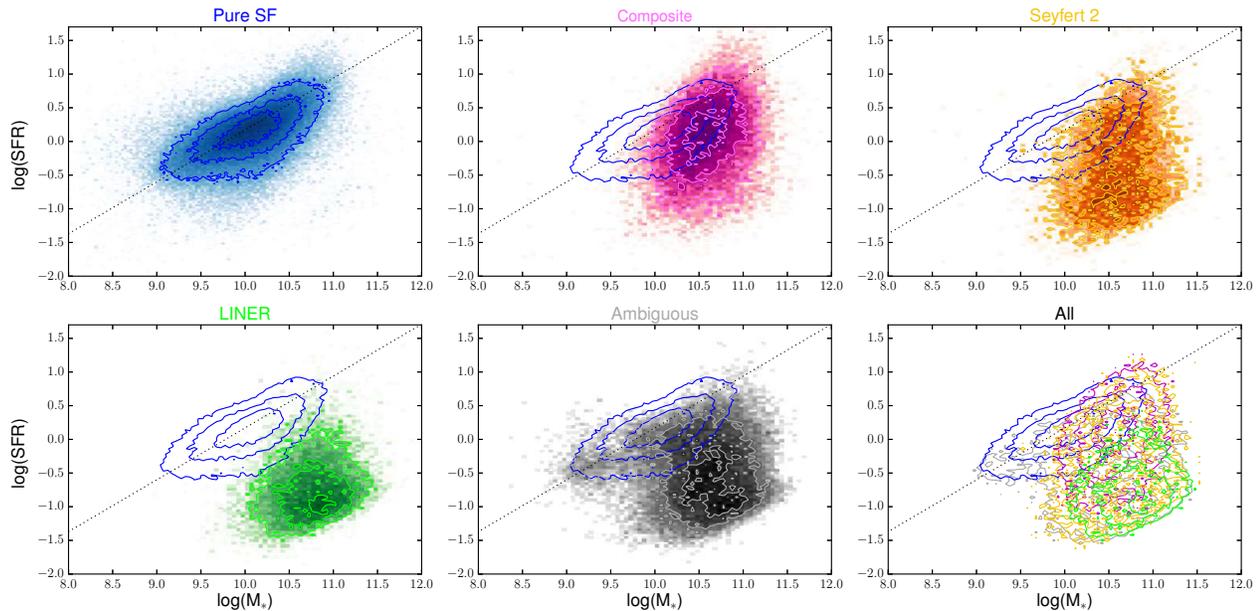}
\caption{SFR as a function of stellar mass for each class of galaxies. Stellar mass and SFRs are in units of M$_\odot$ and M$_\odot$yr$^{-1}$ respectively throughout this work. The top-left to bottom-right panels include galaxies classified as purely star-forming, composite, Seyfert 2, LINER, or Ambiguous, and the final panel contains all classes together. A black dashed line in each panel represents the local MS relation for blue SDSS galaxies determined by \cite{elbaz07}. Contours and colours represent the number density of galaxies in a single class only. Dividing the SFR-M$_*$ space into 150x150 bins, contours are drawn at 10, 30, and 60\% of the maximum number density. The blue contours of the star-forming galaxies are included in all panels to indicate the location of the star-forming MS. Contours from previous panels are also shown in the final panel, which displays the MS of all galaxies.}\label{MS}
\end{figure*}

The canonical main sequence is shown for star-forming galaxies in the first panel of Figure \ref{MS}. The local SDSS relation derived by \cite{elbaz07} from galaxies with blue optical colours is represented by the black dashed line in all panels of Figure \ref{MS}. The SF galaxies form a relationship between log($M_*$) and log(SFR), with most galaxies classified as SF lying along the MS relation of \cite{elbaz07}.

Unlike the pure SF MS, there is almost no relation between stellar mass and SFR for composite galaxies, other than the fact that composite galaxies all have high stellar masses $M_*>10^{10} M_\odot$. This stellar mass limit may also exist for quasar hosts \citep{matsuoka14}. Composite line ratios may be caused by a combination of star formation and AGN activity \citep{yuan10}, a combination of star formation and shock excitation \citep{allen08,rich11}, or a combination of all three processes \citep{leslie14}.
The contours of composite galaxies extend above the top star-forming MS. This region of high SFR could correspond to galaxies which are undergoing intense bursts of star formation \citep{schiminovich07}.
The composite sequence extends from the massive end of the main sequence down to the region where red galaxies, with little active star formation lie. The mean stellar mass of composite galaxies is higher than that of pure SF galaxies, however the mean SFR is lower. This could imply that suppression of star formation has commenced by the composite stage. 

Further bridging the gap between star-forming galaxies and quiescent galaxies are the Seyfert 2s. The majority of Seyfert 2s lie in the quiescent region of the diagram.

LINERs do not fall on the star-forming MS, but lie below it. Most LINERs in our sample clearly contain very little current star formation activity. 
On average, LINERs have a higher stellar mass than Seyfert galaxies (median log($M_*$) of 10.74 and 10.58 $M_\odot$ respectively), consistent with the findings of \cite{kewley06}. 
The strongest difference between Seyferts and LINERs is their Eddington ratios. Nuclear LINERs that contain AGN are found to accrete less efficiently than Seyfert AGN \citep{kewley06}. \cite{kewley06} also found that LINERs are generally found in galaxies with an older stellar population than Seyfert hosts.

LINER-like emission may also be caused by shocks \citep{kewley01}. Shocks can be produced by AGN- related processes (e.g. \citealt{davis12,alatalo14a,alatalo14b}) or in starburst driven winds (e.g\citealt{sharp10,rich11,ho14}). Winds driven by a powerful starburst, may also drive gas out of a galaxy disc and thus suppress the SFR.

The final panel of Figure \ref{MS} shows all the galaxy classes together, with contours at 10, 30, and 60\% of the maximum number density for each class of galaxies. In general composite galaxies have a higher SFR and lower stellar mass than Seyfert2s, which in turn have a higher SFR and lower stellar mass than LINERs, with ambiguous galaxies having the lowest SFRs (see Table 1). As we will discuss in Section 4, this is consistent with an evolutionary sequence through which galaxies evolve from main sequence star-forming galaxies to being quiescent galaxies through the composite, Seyfert and/or LINER classes.

\begin{table}
\caption{For each class: the percentage of the total sample (202708 galaxies), median signal to noise in the H$\alpha$ emission, median (log) stellar masses and star formation rates and their corresponding inter-quartile range.}
\centering
\begin{tabular}{ccccccc}
\toprule
Class & \% & $S/N$ &\multicolumn{2}{c}{log($M_*$)}  & \multicolumn{2}{c}{log($SFR$)}\\ 
\midrule
SF & 60.4& 24 &10.02 & 0.65 & 0.09 & 0.51\\
Composite& 14& 35&10.56& 0.44 & 0.04& 0.70\\
Seyfert 2& 4.1&15 & 10.58 & 0.51 & -0.45& 0.93\\
LINER& 6.5 & 5 &10.74 & 0.46 & -0.79& 0.53\\
Ambiguous& 17.0& 6 &10.65 & 0.64 & -0.53 & 0.79\\
\bottomrule
\end{tabular}
\end{table}

\begin{figure*}
\centerline{\includegraphics[width=0.95\linewidth]{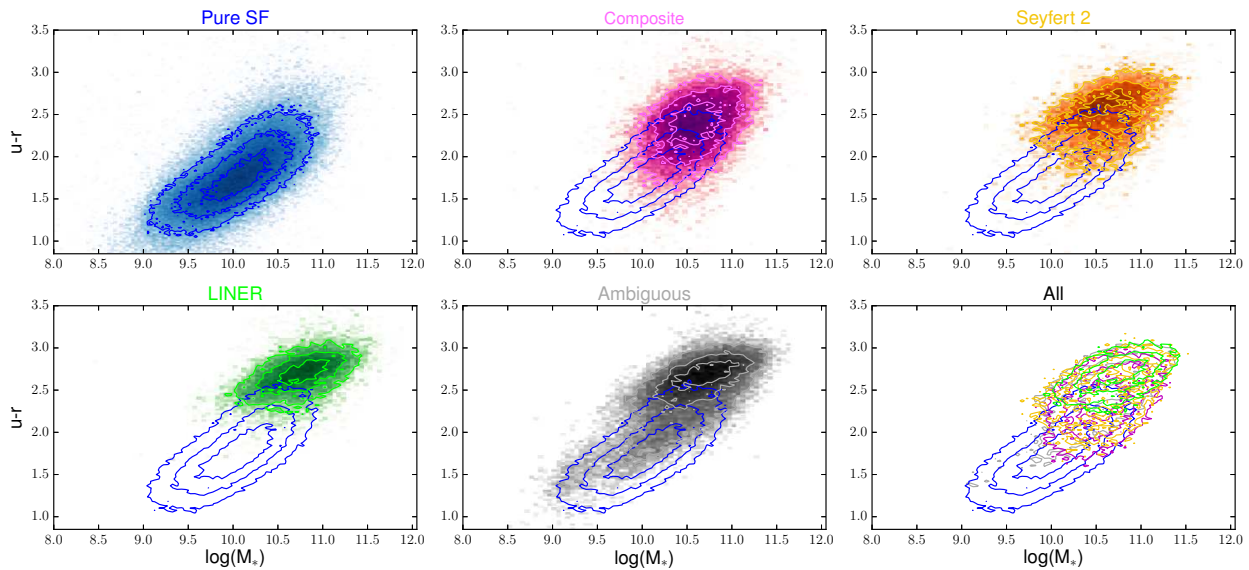}}
\caption{Galaxy reddening-corrected rest-frame colour as a function of stellar mass for each class. The red contour in the Pure SF panel corresponds to the same galaxies in the red contour of Figure 1. Contours are drawn at 10, 30 and 60\% of the maximum density of each class of galaxy.}\label{colour}
\end{figure*}

The colour-mass diagram in Figure \ref{colour}, showing u-r as a function of log($M_*$) reinforces the story  of the MS in Figure \ref{MS}. Purely star-forming galaxies have blue colours and span a range of stellar masses. Composite galaxies have redder colours and are more massive than purely star-forming galaxies. On average Seyfert 2s have redder colours than composite galaxies and are more massive, however there is a significant amount of overlap between the two classes. \cite{schawinski07} found similar results for a sample of early type galaxies and suggested that the overlap implies that galaxies may transition between the Seyfert and composite regions. 
\cite{schawinski07} \& \cite{smolcic09} report a similar trend in the colour-mass diagram,
 with a sequence existing between the blue cloud and the red sequence; from blue star-forming galaxies at the lowest mass, through Seyferts in the green valley, to LINERs at the highest masses and reddest colours.  

\section{Evolutionary Pathways}
We interpret the trends seen in Figures \ref{MS} and \ref{colour} as evidence that optical emission line classes provide important pathways for galaxies reaching the red sequence. Galaxies start their lives on the main sequence forming stars and gaining stellar mass at a steady rate until they reach $M_*\sim 10^{10} M_\odot$. The transition away from the main sequence for our sample only occurs at stellar masses $>10^{10}$M$_\odot$. At these high stellar masses mass-quenching is expected to dominate over environmental quenching, which could involve strangulation, ram-pressure stripping or harassment \citep{peng10}. Our work indicates that galaxies may transition from star-forming through one of, or a combination of composite, Seyfert, and/or LINER stages before achieving quiescence. \cite{karim11, whitaker14} \& \cite{lee15} report a turnover in the MS at stellar masses $\sim10^{10}$. The inclusion of composite and AGN galaxies in these high redshift studies may be the major factor causing the flattening of the main sequence at stellar masses $> 10^{10}$M$_\odot$.

The loci of SFR vs M$_*$ for the composite, Seyfert, and LINER classes are almost perpendicular to the star formation main sequence showing decreasing sSFR (specific star formation rate = SFR/M$_*$) for each class consistent with recent results from \cite{shimizu15} and with seminal results from \cite{salim07}. The SFRs of composite galaxies lie above and below the MS relation. The composite galaxies above the MS are likely starburst galaxies similar to those in \cite{schiminovich07}. Quenching may have commenced in composite galaxies below the MS. Composite galaxy quenching could be tied to large scale gas outflows from winds driven by bursts of star formation, with the composite spectra resulting from shocks embedded in the outflows. Alternatively, the quenching could be tied to AGN activity with the composite spectra resulting from a mixture of star formation and AGN emission. AGN jets can drive a wind which is able to blow out gas from a galaxy. Activity from an AGN can also heat gas, preventing star formation.
Galaxies may evolve off the MS towards the red-sequence, via one or more of the composite, Seyfert and LINER stages.

We emphasize that different phenomena may produce the same emission line ratios for galaxies within each class. Composite emission spectra are not only produced by SF and AGN, but are often shock excited. LINER-like emission may be caused by post AGB stars \citep{singh13,cidfernandes09} or shocks, as well as by low-luminosity AGN \citep{ho93}. To elucidate the nature of galaxies in these classes, and their relationship to star formation quenching, our future work includes wide integral field spectroscopy of statistically significant datasets in each class.

\vspace{-0.7cm}
\section{Conclusion}
Using star formation rates and stellar masses from the MPA/JHU measurements of SDSS galaxies, we have included composite and AGN host galaxies on the local star-forming main sequence and colour-mass diagrams. Our results suggest that:
\begin{itemize}
\item AGN activity (composites, Seyferts, LINERs) appears to play a very important role in ``quenching'' star formation in massive ($M_* >10^{10} M_\odot$) galaxies;
\item the loci of $SFR\ vs.$ $M_*$ for composites, Seyferts and LINERS are nearly perpendicular to the SF MS.  Composites show the broadest range in SFR (+0.8 to -1.8 dex above/below the MS), while most Seyferts lie below the MS ($SFR \sim$ 0 to -2 dex), and LINERs lower still ( $SFR \sim$ -1 to -2 dex); 
\item composites, Seyferts and LINERs plausibly form a set of evolutionary pathways by which massive SF MS galaxies transition into a ``red and dead'' quiescent phase.
\end{itemize}
Future work with large spatially resolved integral field unit data sets, such as those provided by MaNGA \citep{bundy15} and SAMI \citep{croom12}, will likely shed more light on this sequence, by allowing the study of the spatial distribution of star formation in galaxies as a function of galaxy properties such as morphology, stellar mass, BPT classification, and environment.

\section*{Acknowledgements}
We thank the referee for his valuable comments and suggestions. Funding for the
SDSS and SDSS-II has been provided by the Alfred P. Sloan Foundation, the
Participating Institutions, the National Science Foundation, the U.S. Department
of Energy, the National Aeronautics and Space Administration, the Japanese
Monbukagakusho, the Max Planck Society, and the Higher Education Funding
Council for England. The SDSS Web Site is http://www.sdss.org/.
D. S. would like to thank the Distinguished Visitor Program at RSAA, ANU for their generous support while in residence at the Mount Stromlo Observatory, ACT. L. K. is supported by an ARC Discovery Project DP130103925 and gratefully acknowledges an ARC Future Fellowship.

\bibliographystyle{aa}
\bibliography{bibfile}{}
\end{document}